\documentclass[twocolumn,aps,showpacs,floats]{revtex4}
\usepackage{calc}
\usepackage{graphicx}
\begin{document}
\def \beq{\begin{equation}}
\def \eeq{\end{equation}}

\begin{abstract}
We systematically discuss candidate wave functions for the ground state
of the bilayer $\nu = 1$ as the distance between the layers is varied.
Those that describe increased intralayer correlations at finite distance
show a departure from the superflid description for smaller distances.
They may support finite energy meron excitations and  a dissipative
collective mode in the place of the Goldstone mode of the ordered phase
i.e. describe a vortex metal phase, or imply even an incompressible,
pseudospin liquid, behavior. Therefore they describe possible outcomes of
quantum disordering at finite distance between the layers. The vortex metal phase
may show up in experiments in the presence of disorder at lower temperatures
and explain the observed ``imperfect superfluidity", and the pseudospin liquid
phase may be the cause of the thermally activated (gapped) behavior of
the longitudinal and Hall resistances at higher temperatures in counterflow
experiments.
\end{abstract}


\title{Wavefunctional approach to the bilayer $\nu = 1$ system and a possibility for
a double non-chiral pseudospin liquid}
\author{M.V. Milovanovi\'{c}}
\affiliation{Institute of Physics, P.O.Box 68, 11080 Belgrade,
Serbia}

\date{\today}
\maketitle \vskip2pc
\narrowtext
\section{Introduction}
The bilayer $\nu = 1$ quantum Hall (QH) system consists of two layers of 2D electron gases,
each with a filling factor $1/2$, that are brought together at distance
comparable to the average distance between electrons inside layers and
the tunneling is negligible. The physics of the bilayer has been the focus
of much of experimental and theoretical work. We will mention only two
major experimental findings. First in the experiment of Spielman et al. \cite{spi}
 there was a very pronounced (``spectacular") zero-bias peak in the
tunneling conductance between layers. Second in the experiments of Kellogg et al. \cite{kel}
and Tutuc et al. \cite{tut},
 in the counterflow setting of the bilayer (where in each layer the
current is of opposite sign than the other) the resulting longitudinal
and Hall resistances were dropping to zero in the limit of zero temperature.
Both types of experiments signal a superfluid behavior in the pseudospin (relative
between layers) channel of the system. An elaborated theoretical work
of Moon et al. \cite{moo} described this behavior in the framework of an easy-plane
ferromagnet (where the layer index represents the electron pseudospin
degree of freedom) and ensuing XY model description. Therefore the
superfluid behavior was expected. Usually it is described as a consequence
of an excitonic condensate inside the bilayer where each exciton
stems from the electron coupling to its correlation hole just opposite
in the other layer. Nevertheless some discrepancies were noticed. In the
first experiment of Spielman et al. the peak due to some unknown
source of dissipation, is not as high and narrow as in the analog,
Josephson  tunneling experiments in superconductors. In the
second experiment of Kellogg et al. and Tutuc et al. there were no Kosterlitz-Thouless
transition signatures in the measured counterflow resistance. It seems
merons (vortices) are liberated (due to some unknown cause) from their
confinement in the expected superfluid down to very low temperatures.

These findings point out that a careful investigation of quantum
fluctuations or quantum disordering at finite $d$ (distance between
the layers) is necessary. One way to approach this question is to adopt
the Laughlin approach to the fractional QH effect and look for approximate but very
close to the ground states wave functions at finite $d$. That this is
possible it was numerically demonstrated in Ref. \cite{srm} for not large $d$.
The functions that approximate the true ground states incorporate
the effects of quantum disordering by allowing the presence of
composite fermions (CFs) next to composite bosons (CBs) , another transformed
electrons, characterizing the bose condensate of the assumed superfluid.
CFs connect in a special way to the condensate to ensure and maintain
the rigidity of the bose condensate.

In this work we will systematically discuss candidates that we expect
would approximate very well the ground state wave functions at finite $d$.
Some of them we expect would be in a competition with the ones that are
 ferromagnetically ordered and possess a Goldstone mode. Among them we will point out to one
that describes an incompressible (in all channels) state. If the state were
a true ground state, it would be a ground state of topological phase
that supports quasiparticles - merons with finite excitation energy. In
fact this would be a realization of what is usually called a spin liquid
phase with two kinds of semionic quasiparticles - i.e.
a double spin liquid, with the only difference that here we work
with pseudospin instead of spin. We will
use interchangeably words double pseudospin liquid
and pseudospin liquid, for short, denoting the same phase.
 This gapped phase might be the one
that appears in counterflow experiments causing activated
(gapped) behavior of the longitudinal and Hall resistances for a range of
higher temperatures.
Another competing possibility with possibly finite excitation energy for merons is
a compressible version in which quantum disordering has allowed
weakly coupled meron-antimeron pairs i.e. a vortex metal phase. This
state supports a dissipative collective mode that comes in the place
of the Goldstone mode of the ordered phase. The physics of this state
should be relevant for the explanation of the dissipation effects in
the experiments, in which disorder, at lower temperatures, would
dissociate the expected closed loops of meron-antimeron pairs in the
topological phase.

The paper is organized as follows: Sections II, III, and IV contain
results of our paper, and Sections V and VI discussion and conclusions.
Section II is an introduction to the wavefunctional approach to the
bilayer, Section III in two different approaches with compatible
conclusions describes the physics of a candidate wave function -
vortex metal state, and Section IV introduces its modification due
to CF pairing - a pseudospin liquid state.

\section{Wavefunctional Approach to the Bilayer $\nu = 1$ system}
 If the distance between the layers is of the order or less the magnetic
length - average distance between electrons inside the layers,
interlayer Coulomb interaction will force the system, at total
filling factor one and in the conditions of QH effect, to form the (111) state.
The state is a simple generalization of the single-layer
filling-factor-one case i.e. completely filled lowest Landau level (LLL)
of one species electrons, precisely it is
\begin{equation}
\Psi_{111}(z_{\uparrow}, z_{\downarrow}) =
\prod_{i<j}(z_{i\uparrow}-z_{j\uparrow})
\prod_{k<l}(z_{k\downarrow}-z_{l\downarrow})\prod_{p,q}(z_{p\uparrow}-z_{q\downarrow})
\end{equation}
where $z_{i \uparrow}$ and $z_{i \downarrow}$ are two-dimensional
complex coordinates of electrons in upper and lower layer respectively and
we omitted the Gaussian factors. If layers are far apart
each one will be a separate system at filling factor one-half.
Numerous studies \cite{rr} show that the ground state at that filling factor
is a generalization of the Laughlin construction which
includes a single Slater determinant of noninteracting particles
in zero magnetic field, i.e.
\begin{equation}
\Psi_{1/2}(w) = {\cal P} \{\Phi_{f}(w,\overline{w})\prod_{i<j}(w_{i\uparrow}-w_{j\uparrow})^2\}
\end{equation}
where $\Phi_{f}$ is the determinant and ${\cal P}$ represents projection to LLL.
In Refs. \cite{rr,hlr} it was shown that relevant, underlying
composite particles in this case are CFs. On the other hand the
relevant, weakly interacting composite particles in the (111) case
are CBs \cite{zha,ismvm}.

Now let us start from (111) case, increase the distance and
introduce one-half correlations in the (111) state in a minimal way,
preserving (111) intercorrelations of newly introduced CFs with
all other remaining CBs. This means that though we are perturbing
(111) state, we are assuming its inherent rigidity.
The wave function that describes this is

\begin{eqnarray}
 \Psi_{bbf}=\cal{PA} &\{&
\prod_{i<j}(z_{i\uparrow}-z_{j\uparrow})
\prod_{k<l}(z_{k\downarrow}-z_{l\downarrow})\prod_{p,q}(z_{p\uparrow}-z_{q\downarrow}) \nonumber\\
&& \times \;
\Phi_{f}(w_{\uparrow},\overline{w}_{\uparrow})\prod_{i<j}(w_{i\uparrow}-w_{j\uparrow})^2
\nonumber \\
&& \times \; \Phi_{f}(w_{\downarrow},\overline{w}_{\downarrow})\prod_{k<l}(w_{k\downarrow}-w_{l\downarrow})^2 \nonumber \\
&& \times
\prod_{i,j}(z_{i\uparrow}-w_{j\uparrow})\prod_{k,l}(z_{k\uparrow}-w_{l\downarrow}) \nonumber \\
&& \times \prod_{p,q}(z_{i\downarrow}-w_{q\uparrow})\prod_{m,n}(z_{m\downarrow}-w_{n\downarrow})
  \}
\label{first}
\end{eqnarray}
where we omitted the Gaussian factors, $\cal{P}$ denotes the
projection to the LLL, $\cal{A}$ - antisymmetrization between bose and
fermi variables in each layer separately, and $\Phi_{f}$s are
Slater determinants of free waves. In the thermodynamic limit the
flux - number of particle relationship is
\begin{eqnarray}
\label{br}
N_{\Phi}&=&
N_{b\uparrow}+N_{b\downarrow}+N_{f\uparrow}+N_{f\downarrow} \nonumber\\
&=& 2N_{f\uparrow}+N_{b\uparrow}+N_{b\downarrow} \\
&=& 2N_{f\downarrow}+N_{b\uparrow}+N_{b\downarrow}\nonumber.
\end{eqnarray}
Consequently we must have $N_{f\uparrow}=N_{f\downarrow}$ but
there is no constraint on the relative number of bosons.

In fact, once we adopt the hypothesis that with increasing $d$ (distance)
CFs are slowly nucleating we are left, because of the flux-counting
arguments, with only two (simple, ala Laughlin) possibilities for
interpolating ground state.
The second possibility has instead of the last two lines in
(\ref{first}) for the intercorrleations between the bose and
fermi parts, the following expression,
\begin{equation}
\label{izm}
\prod_{i,j}(z_{i\uparrow}-w_{j\uparrow})^2
\prod_{k,l}(z_{k\downarrow}-w_{l\downarrow})^2.
\end{equation}
These intercorrelations are of the fermi part type, i.e. they are
intralayer correlations, a consequence of possibly more important
intralayer Coulomb interactions.

In this case we have
\begin{eqnarray}
N_{\Phi}&=& 2N_{f\uparrow}+2N_{b\uparrow} \nonumber \\
&=& 2N_{f\downarrow}+2N_{b\downarrow} \nonumber \\
&=& 2N_{f\uparrow}+N_{b\uparrow}+N_{b\downarrow} \\
&=& 2N_{f\downarrow}+N_{b\uparrow}+N_{b\downarrow} \nonumber
\end{eqnarray}
i.e. both bose and fermi numbers are constrained to
$N_{f\uparrow}=N_{f\downarrow}$ and
$N_{b\uparrow}=N_{b\downarrow}$.

In the small particle (5+5) numerical study in \cite{srm} the first
possibility had much larger overlaps, presented in \cite{srm}, with the true
ground state as it was evolving with distance (and the ratio
between bosons and fermions was changing) than the second
possibility (Eq.(\ref{izm})). The overlaps of the first possibility were slowly
decreasing with distance so that no definite conclusions can be
drawn about the precise evolution of the system near the
transition region. But for smaller $d$ the relevance of the first
possibility for the evolution was established.

We will not discuss the states that we naturally expect would
interpolate between these two limits, small $d$ (Eq.(\ref{first}))
and near the transition (Eq.(\ref{izm})) limit. In these states
some of the fermions would connect via bose (111) type
intercorrelations to the bose part and some via fermi type.

It is no wonder that the first possibility is more relevant for
smaller $d$. It also allows the imbalanced ($N_{1}\neq N_{2}$)
situation, which is required due to the theoretical and
experimental insight gained \cite{wen,spi,moo,gir} about
the existence of the Goldstone mode connected with the particle
number difference. (The projected to definite particle number
picture, which we here discuss, must incorporate that
physics.) Later we will show that the state is incompressible
(in the charge channel), at
least in the scope of a suggested Chern-Simons picture.

\subsection{Chern-Simons approach}
 In this section we will employ the classical CS approach,
described in \cite{zha} and \cite{hlr} with its usual RPA (random phase
approximation) to find out the linear response of the system that
supports  the ground state in Eq.(\ref{first}). This approach is advanced in
\cite{sm} but at the same time there, a conclusion is drawn that
the classical CS approach in the RPA allows us with not much work
to find out qualitatively the physics of the linear response. Here
we have still another problem; there is no obvious way to
implement antisymmetrization procedure of the wave function in
Eq.(\ref{first}). The antisymmetry of electronic wave functions
comes naturally and automatically only in only CB or only CF
theories. Therefore to simplify matters from the start we choose
the classical CS approach and neglect the antisymmetrization
requirement.

Without the antisymmetrization the wave function in (\ref{first})
would be the ground state (in the RPA and when the projection to
LLL is neglected) of the following CS theory
\begin{eqnarray}
\label{lagr}
\mathcal{L}&=&\sum_{\sigma}\{i\Psi^\dagger_{\sigma}(\partial_{0}+ia^{\sigma}_{0}-iA_{0}-i\sigma
B_{0})\Psi_{\sigma} \nonumber \\
&+&
\frac{1}{2m}\sum_{k}\Psi^\dagger_{\sigma}(\partial_{k}+ia^{\sigma}_{k}-iA_{k}-i\sigma
B_{k})^2\Psi_{\sigma} \} \nonumber \\
&+&\sum_{\sigma}\{i\Phi^\dagger_{\sigma}(\partial_{0}+ia_{0}-iA_{0}-i\sigma
B_{0})\Phi_{\sigma}  \nonumber \\ &+&
\frac{1}{2m}\sum_{k}\Phi^\dagger_{\sigma}(\partial_{k}+ia_{k}-iA_{k}-i\sigma
B_{k})^2\Phi_{\sigma} \} \nonumber \\
&+& \sum_{\sigma} \frac{1}{2\pi}\frac{1}{2}
a^{\sigma}_{0}(\vec{\nabla}\times \vec{a}^{\sigma}) \nonumber \\
&-&\frac{1}{2}\sum_{\sigma}\int d\vec{r}'
\delta\rho_{\sigma}(\vec{r})V_{a}(\vec{r}-\vec{r}')\delta\rho_{\sigma}(\vec{r}') \\
&-&
\int d\vec{r}'
\delta\rho_{\uparrow}(\vec{r})V_{e}(\vec{r}-\vec{r}')\delta\rho_{\downarrow}(\vec{r}')
\nonumber
\end{eqnarray}
where $\sigma$ is the layer index, taking $\uparrow$ and $\downarrow$ values as a
variable, $\Psi_{\sigma}$ and $\Phi_{\sigma}$, are CF and CB
$\sigma$-layer fields respectively, and $V_a$ and $V_e$ are intra
and inter-Coulomb interactions. $\delta\rho_{\sigma}$'s are given
as sums of CBs and CFs, i.e.
$\delta\rho_{\sigma}=\delta\rho_{\sigma}^{F}+\delta\rho_{\sigma}^{B}$,
as the most natural choice for the electron density in this
distinguishable picture. External fields, $A$ and $B$, couple to
charge and pseudospin (up minus down) degrees of freedom
respectively. The CS, gauge fields, $a_{\uparrow}$ and
$a_{\downarrow}$, are related to $a$ as
$$
a=\frac{a_{\uparrow}+a_{\downarrow}}{2},
$$
and in this way, in the mean field approximation, reproduce the
relations encoded in the ground state given by Eq.(\ref{first}).

The antisymmetrization can be implemented by the following
constraint,
\begin{equation}
\label{veza}
\vec{S}\cdot
\vec{S}=\frac{N_{\sigma}}{2}(\frac{N_{\sigma}}{2}+1),
\end{equation}
for each layer separately so that $N_{\sigma}$ denotes the number
of electrons in the layer. $\vec{S}$ denotes a generalized spin of
the layer obtained by integrating over the volume of the system of
the following field density,
$$
\hat{\Psi}^\dagger_{\sigma}(\vec{r})\frac{\vec{\sigma}}{2}\hat{\Psi}_{\sigma}(\vec{r}).
$$
$\hat{\Psi}(\vec{r})$ denotes a spinor for which we have,
$$
\hat{\Psi}_{\sigma}(\vec{r})=\left[ {\begin{array}{c}
                                        U^{\dagger}_{b}\Phi_{\sigma}
                                        \\[2mm]
                                        U^{\dagger}_{f,\sigma}\Psi_{\sigma}
                                     \end{array}} \right]
$$
where $U_{b}$ and $U_{f,\sigma}$ are the CS unitary
transformations, i.e.
\begin{eqnarray}
&\;\;\;\;\;\;\;\;\;\;&  U_{b}(r)= \nonumber \\
&\;\;&  exp \ \bigg\{i\int d^2r'
arg(r-r')(\rho_{b\sigma}(r')+\rho_{b-\sigma}(r') \nonumber \\
&\;\;\;\;\;\;\;\;\;\;& \;\;\;\;\;\;\;\;+\rho_{f\sigma}(r')+\rho_{f-\sigma}(r'))\bigg\} ,
\end{eqnarray}
and
\begin{eqnarray}
&&U_{f\sigma}(r)= \nonumber \\
&& exp \ \bigg\{i\int d^2r'
arg(r-r')(2\rho_{f\sigma}(r')+ \nonumber \\
&& \;\;\;\;\;\;\;\;\rho_{b\sigma}(r')+\rho_{b-\sigma}(r'))\bigg\},
\end{eqnarray}
$\Phi_{\sigma}$ and $\Psi_{\sigma}$, are already introduced bose
and fermi field respectively, and $\vec{\sigma}$ are the usual
Pauli matrices. The idea behind the constraint in Eq.(\ref{veza}) is
simple; it uses the fact that spin $\frac{1}{2}$ particles,
$N_{\sigma}$ of them, must necessarily, in the states for which the constraint in
 Eq.(\ref{veza}) is true i.e. have $\vec{S}\cdot\vec{S}$ at its
largest possible value, be completely symmetric in the spin space
and necessarily antisymmetric in the real space.

We still have to fix $S_{z}$, which is the second constraint, i.e.
the number difference between CBs and CFs in the layer, to project
the artificially introduced spin $\frac{1}{2}$ problem to the
reduced Hilbert space and our problem.

What we can immediately notice is that the constraint in Eq.(\ref{veza})
will introduce  the terms that interchange bosons and
fermions at different points instantaneously which conforms to our
idea of indistinguishability. Although it is easy to formulate it
is very hard to implement the constraint. Only, maybe, if we have
overwhelming number of bosons or fermions we will be allowed to
neglect the constraint as it is usual in the case of bose fluids
with the natural decomposition into classical-macroscopic,
condensate part and normal part.

Nevertheless there is a much deeper reason that allows us to neglect
the antisymmetrization requirement. The reason is that, just like
in a hierarchical construction and as it will be much more
clear later, CFs represent meron excitations (meron-antimeron pairs,
see Appendix) that quantum disorder the (111) state. As it is
usual when we discuss the dual picture of the FQHE \cite{blo},
we do not extend the antisymmetrization requirement to the
quasiparticle part of the electron fluid.

We want to prove that indeed the static density-density response
that follows from the Lagrangian in Eq.(\ref{lagr}) shows the
incompressibility of the system with the ground state in Eq.(\ref{first})
in which the antisymmetrization is neglected. The charge of the
system is related to $\delta\vec{a}$ as
$$
\delta\rho=\delta\rho_{\uparrow}+\delta\rho_{\downarrow}=\frac{ik\delta
a}{2\pi}
$$
where, as we work in the transverse gauge, $|\delta\vec{a}|=\delta
a$ is the transverse space component of the vector. For the
response calculations (with $B_{0}=B=0$) we will adopt the
conventions introduced in \cite{hlr}. Integrating out the quadratic
terms in fermionic fields to the second order in gauge fields
(RPA) \cite{hlr}, then introducing density-angle variable for the
bosonic fields and expanding the action in them neglecting the
amplitude-density derivatives \cite{zha}, we get
\begin{eqnarray*}
\mathcal{L}_{eff}&=&\sum_{\sigma}\frac{1}{2}[K_{00}^{\sigma}(\delta
a_{0}^{\sigma})^2+ K_{11}^{\sigma} (\delta a^{\sigma})^2] \nonumber \\
&+&
\frac{1}{2\pi}\frac{1}{2}\sum_{\sigma}a^{\sigma}_{0}(-k)ika^{\sigma}(k)
\nonumber \\
&+&\sum_{\sigma}\delta\rho^{B \sigma}(-k)(i\omega
\Theta_{\sigma}(k)-\delta
a^{\sigma}_{0}(k)) \nonumber \\
&+&\sum_{\sigma}(-)\frac{\overline{\rho}_{b}}{2m}[k^2\Theta^2_{\sigma}+(\delta
a)^2] \nonumber \\
&-& \frac{1}{2}\frac{k^2(\delta
a)^2}{(2\pi)^2}V_{c}(k)-\frac{1}{2}V_{s}(k)|\delta\rho^{B \uparrow}-\delta\rho^{B \downarrow}+\frac{1}{2\pi}ika^{s}|^2,
\end{eqnarray*}
where $V_{c}(k)=\frac{V_{a}(k)+V_{c}(k)}{2}$,
$V_{s}(k)=\frac{V_{a}(k)-V_{e}(k)}{2}$,
$K_{00}=K_{00}^{\uparrow}=K_{00}^{\downarrow}$ and
$K_{11}=K_{11}^{\uparrow}=K_{11}^{\downarrow}$ are density-density
and current-current response functions of a free fermionic sustem
respectively \cite{hlr}, $\Theta_{\uparrow}$, $\Theta_{\downarrow}$ are
bosonic angle variables and $\overline{\rho}_{b}$ averaged bosonic
density per layer. In the static ($\omega=0$) case we are left
with the following charge part, decoupled from the pseudospin part
in the RPA,
\begin{eqnarray*}
\mathcal{L}_{eff}&=&K_{00}(\delta a_{0})^2+K_{11}(\delta
a)^2+\frac{1}{2\pi}a_{0}ika \\
&+& \delta\rho_{c}^{b}\delta
a_{0}-\frac{2\overline{\rho}_{b}}{2m}|\delta
a|^2-\frac{1}{2}\frac{k^2(\delta a)^2}{(2\pi)^2}V_{c}(k).
\end{eqnarray*}
The integration over $\delta\rho_{c}^{b}$ gives the constraint
$a_{0}=A_{0}$ and consequently the integration over $\delta a$
gives for the (static) density-density response,
$$
\frac{(\frac{1}{2\pi})^{2} k^2}{\frac{2\overline{\rho}_{b}}{m}+\frac{k^2V_{c}(k)}{(2\pi)^2}-2K_{11}},
$$
where $K_{11}=-\frac{k^2}{12\pi m}$ \cite{hlr}. As long as there is a
finite density of bosons ($\overline{\rho}_{b}$) the system is
quantum Hall, incompressible (and the above expression vanishes in
$k\rightarrow 0$ limit). Even the presence of composite fermions
i.e. particles which we would naively consider to represent
gapless degrees of freedom in the system, does not lead to
compressibility. A small fluctuation in the charge density
($\delta\rho_{f}=\delta\rho_{f}^{\uparrow}+\delta\rho_{f}^{\downarrow}$)
of the composite fermions lead to a fluctuation in the bosonic
charge density for which we know we need a finite amount of energy
to create \cite{zha}.

\section{A State with Possibly Deconfined Merons}
The second possibility, state
\begin{eqnarray}
\Psi_{bff} = {\cal P A} \{& \prod_{i<j}(z_{i \uparrow} - z_{j \uparrow})
                            \prod_{k<l}(z_{k \downarrow} - z_{l \downarrow}) & \nonumber \\
                           & \prod_{p,q}(z_{p \uparrow} - z_{\downarrow}) & \nonumber \\
& \Phi_{f}(w_{\uparrow},\overline{w}_{\uparrow}) \prod_{i<j} (w_{i \uparrow} - w_{j \uparrow})^{2} & \nonumber \\
&\Phi_{f}(w_{\downarrow},\overline{w}_{\downarrow}) \prod_{i<j} (w_{i \downarrow} - w_{j \downarrow})^{2}
& \nonumber \\
\label{second}
&\prod_{i,j}(z_{i \uparrow} - w_{j \uparrow})^{2}  \prod_{k,l}(z_{k \downarrow} - w_{l \downarrow})^{2}\}   &
\end{eqnarray}
should be relevant for the transition into two decoupled Fermi seas region. In
the following section we will explore its properties; in the first part we will show
that in this state possibly a finite energy is needed to excite meron, and in the second part
we will show that the state is compressible in the pseudospin channel and nearly supports
a gapless pseudospin mode. The first property tells us that in this state merons may be
deconfined relative to the (111) and Eq.(\ref{first}) states in which merons as
vortices of the pseudospin superflud are confined. If it were not for the second
property, the state would possibly describe a quantum Hall, topological phase with four kinds
of gapped, meron quasiparticle excitations.
\subsection{The screening of meron}
In the following we will show that the screening charges of the meron excitation in
the plasma analogy of the state in Eq.(\ref{second}) are localized, not of long range,
and may not lead to the usual logarithmic divergence with the size of the system of
the energy required to excite a meron.

Namely we will use an effective expression from Ref.\cite{mvmis} for the meron
excitation that in the (relative change in) density calculations for the meron state
leads to results valid in the longwavelength limit. The effective expression is the
one reduced from the well known expression in the (111) state \cite{moo}, and in the Ref. \cite{mvmis}
it was shown that leads to the logarithmic divergence in the energy to excite meron in
both (111) and the first mixed state (Eq.(\ref{first})) that describe the pseudospin
condensate at small $d$. Therefore the effective expression gives the expected
behavior of a meron excitation in these states. As in the Laughlin quasihole construction,
the meron effective construction multiplies the ground state, $\Psi_{o}$. Explicitly
the construction is
\begin{equation}
\prod_{i} \frac{z_{i \uparrow} - w}{|z_{i \uparrow} - w|} \exp\{ \sum_{i}
\frac{C}{2 |z_{i \uparrow} - w|} \} \Psi_{o},
\label{mcon}
\end{equation}
where $C$ is a constant, found to be equal to $C = 0.80$ in the case of the (111) state.
We emphasize that Eq.(\ref{mcon}) is an effective expression, modeled for the task of
density calculations, and in no way an expression that would be valid in the
short distance limit or stand for a meron construction in the lowest Landau level.

In the plasma language of the Ref.\cite{mvmis}, and as an interpretation of the
squared norm of the construction (Eq.(\ref{mcon})) we have that an impurity at point
$w$ connects via the interaction $ C/|z_{i \uparrow}|$ to the $\uparrow$ particles
of the plasma defined by the state, $\Psi_{o}$. To obtain the charge contributions
($\uparrow$ and $\downarrow$) far away from the center of excitation in the second
state (Eq.(\ref{second})), i.e. if we take in Eq.(\ref{second}) that $\Psi_{o}$ is
the second state, we use the same type of approximations explained in Refs.
\cite{mvmes} and \cite{mvmis}. We assume that the classical partition function
of the classical system that is defined by the square norm of the second state
has the property of screening, and therefore that the essential physics including
the description of the screening charges of impurities can be found by summing
so-called chain diagrams. The contributions to the screening charges of an impurity
 can be easily visualized as a type of chain diagrams shown in Figs. 2 and 3 which connect
impurity $(w)$ to the probing point $(r)$ on the right-hand side. In the Figures
the straight line denotes the $\frac{1}{r}$ interaction, i.e. $\frac{1}{q}$ in
the momentum space by which the meron in Eq.(\ref{mcon}) connects to $\uparrow$
particles, the wriggly line denotes the $\ln\{r\} (\sim \frac{1}{q^{2}})$ interaction
i.e. the Coulomb interaction of 2D plasma, and there are two kinds of vertices.
For bosonic quasiparticles we have a vertex that equals to their densities,
$n_{\uparrow}$ or $n_{\downarrow}$, but for fermionic quasiparticles the value
of the vertex is the static structure factor of free Fermi gas i.e. $s_{\uparrow}(q)$ or
$s_{\downarrow}(q)$ for which we have $s_{\uparrow}(q)=s_{\uparrow}(q)\sim q$ in the
longwavelength limit.
\begin{figure}
\centering
\includegraphics[width=\linewidth]{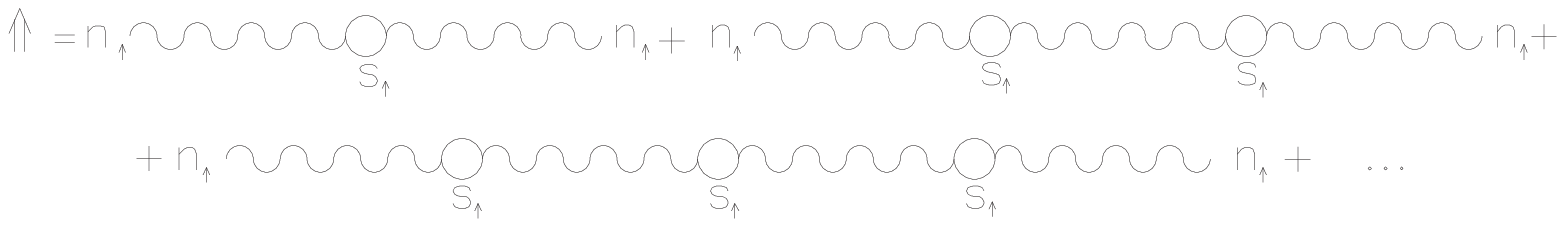}
\caption{The infinite sum of Eq.(\ref{sum})}\label{one}
\end{figure}
\begin{figure}
\centering
\includegraphics[width=\linewidth]{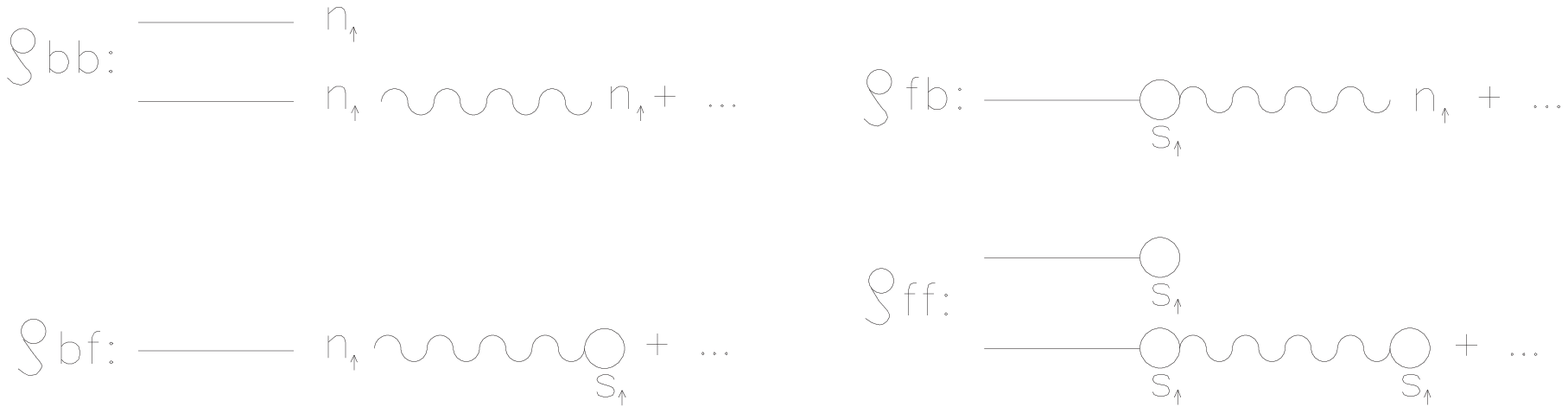}
\caption{The up  density screening}\label{two}
\end{figure}
\begin{figure}
\centering
\includegraphics[width=\linewidth]{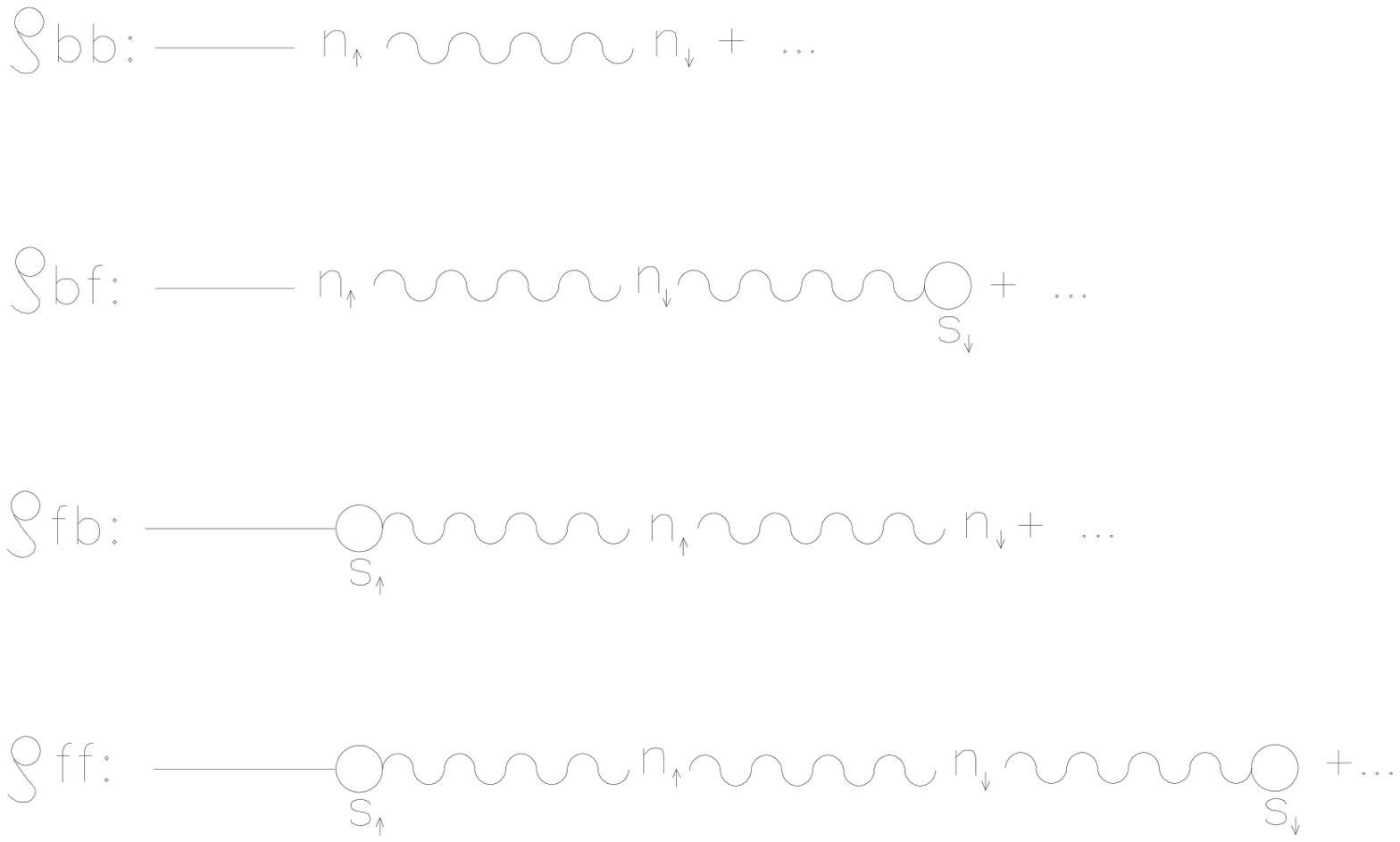}
\caption{The down density screening}\label{three}
\end{figure}
To calculate the screening charges let us introduce two infinite sums;
the one for $\uparrow$ particles is depicted in Fig. 1 and equals
\begin{equation}
\Uparrow = \frac{4 s_{\uparrow} V^{2} n_{\uparrow}^{2}}{1 - 2 s_{\uparrow} V}  \stackrel{q \rightarrow 0}{\longrightarrow}
- 2 V n_{\uparrow}^{2},
\label{sum}
\end{equation}
where $V \sim \frac{1}{q^{2}}$, and similarly we have $\Downarrow$ for  $\downarrow$
particles.

Using these infinite sums we are able to write the effective interaction
between $n_{\uparrow}$  vertices in a compact form,
\begin{eqnarray}
{\cal V}(n_{\uparrow}, n_{\uparrow}) =
& \frac{n_{\uparrow} V n_{\uparrow}}{1 - V(n + \Uparrow + \Downarrow)} + \frac{\Uparrow}{1 - \frac{\Uparrow}{n_{\uparrow}}} + & \nonumber \\
& \frac{n_{\uparrow} V \Uparrow + \Uparrow V n_{\uparrow}}{1 - V(n + \Uparrow + \Downarrow)}
\frac{1}{1 - 2 \frac{\Uparrow}{n_{\uparrow}}} \stackrel{q \rightarrow 0}{\longrightarrow} - n_{\uparrow}. &
\label{nunu}
\end{eqnarray}
In Eq.(\ref{nunu}) we also stated the value of the interaction in the
longwavelength limit. The introduction of the infinite sums comes naturally
because of the type of intercorrelations that exist in the electronic
wave function. In this way we can express the effective interaction between a
bosonic and fermionic $\uparrow$ vertex, ${\cal V}(n_{\uparrow},s_{\uparrow})$, as
\begin{eqnarray}
{\cal V}(n_{\uparrow},s_{\uparrow}) = & n_{\uparrow} \frac{2 s_{\uparrow} V}{1 - 2 s_{\uparrow} V} + & \nonumber \\
&{\cal V}(n_{\uparrow}, n_{\downarrow}) \frac{2 s_{\uparrow} V}{1 - 2 s_{\uparrow} V}
\stackrel{q \rightarrow 0}{\longrightarrow} 0, &
\end{eqnarray}
with 0 as the value in the longwavelength limit. Similarly we have for the effective
interaction between fermionic $\uparrow$ vertices,
\begin{eqnarray}
{\cal V}(s_{\uparrow},s_{\uparrow}) = & (s_{\uparrow})^{2} \frac{2 V}{1 - 2 s_{\uparrow} V} + & \nonumber \\
&{\cal V}(s_{\uparrow}, n_{\downarrow}) \frac{2 s_{\uparrow} V}{1 - 2 s_{\uparrow} V}
 \stackrel{q \rightarrow 0}{\longrightarrow}  - s_{\uparrow}(q). &
\label{fl}
\end{eqnarray}
Now we can combine all these effective interaction expressions to find out
the screening charges with up pseudospin of the meron excitation in Eq.(\ref{mcon}).
The first contribution ($\rho_{bb}$ in Fig. 2) is due to the meron connection to $\uparrow$ bose
quasiparticles and, if we denote the direct interaction to $\uparrow$ bose
quasiparticles by $V_{m} \sim \frac{1}{q}$, it is
\begin{equation}
V_{m} + V_{m} \frac{{\cal V}(n_{\uparrow}, n_{\downarrow})}{n_{\uparrow}}
\stackrel{q \rightarrow 0}{\longrightarrow} 0.
\end{equation}
Then the second contribution ($\rho_{bf}$ in Fig. 2) is through the direct interaction to up bose
quasiparticles that ends up with up fermionic quasiparticles; it is
\begin{equation}
V_{m} \frac{{\cal V}(n_{\uparrow}, s_{\downarrow})}{s_{\uparrow}}
\stackrel{q \rightarrow 0}{\longrightarrow}  0.
\end{equation}
The third contribution ($\rho_{fb}$ in Fig. 2) comes from the direct connection to up fermionic quasiparticles
that ends up with up bose quasiparticles. This contribution is similar to the
previous one and has the same limit. The last, fourth contribution ($\rho_{ff}$ in Fig. 2) connects through
up fermi quasiparticles again up fermi quasiparticles, and equals
\begin{equation}
V_{m} + V_{m} \frac{{\cal V}(s_{\uparrow}, s_{\downarrow})}{s_{\uparrow}}.
\end{equation}
This contribution is again equal to zero according to Eq.(\ref{fl}) in the
longwavelength limit.

For the meron screening by down quasiparticles we need ${\cal V}(n_{\uparrow}, n_{\downarrow})$
interaction, and similarly to the same pseudospin interaction we get
\begin{eqnarray}
{\cal V}(n_{\uparrow}, n_{\downarrow}) =
& \frac{n_{\uparrow} V n_{\uparrow}}{1 - V(n + \Uparrow + \Downarrow)}  + & \nonumber \\
& \frac{n_{\uparrow} V \Downarrow + \Uparrow V n_{\downarrow}}{1 - V(n + \Uparrow + \Downarrow)}
\frac{1}{1 - 2 \frac{\otimes}{n_{\sigma}}},  &
\label{nund}
\end{eqnarray}
where $\otimes$ is $\Uparrow$ or $\Downarrow$ and $n_{\sigma}$ is  $n_{\uparrow}$ or $ n_{\downarrow}$ whether
we attach $\otimes$ to the right or left side of a diagram respectively. In this way
we get, in the longwavelenght limit,
\begin{equation}
{\cal V}(n_{\uparrow}, n_{\downarrow}) \stackrel{q \rightarrow 0}{\longrightarrow} 0.
\end{equation}
Because of this any contribution of the down screening charges in which
participate bose quasiparticles is equal to zero. In fact
${\cal V}(n_{\uparrow}, n_{\downarrow})$ participate in each of four possible
contributions, see Fig. 3, leading us to the conclusion that both, $ \rho_{\uparrow}$ and
$\rho_{\downarrow}$, screening charges tend to zero in the
longwavelength limit. Therefore the screening charges are short ranged and
localized just like in the case of any one-component quantum Hall system -
 a topological phase.

A comment is in order here. Although our plasma approach is straightforward
and leads clearly to the localized meron screening charges it does not
automatically lead to an overall finite meron excitation energy because
we can draw a conclusion only for the charging part of the energy
(charge ($\uparrow$ and $\downarrow$) difference squared). But it
certainly signals a possibility for meron deconfinement.
\subsection{The pseudospin degrees of freedom}
Now, in the scope of already introduced Chern-Simons theory, we want to
investigate the pseudospin channel of the state in Eq.(\ref{second}). In
particular we want to know wheather the pseudospin degrees of freedom are
compressible and support a gapless mode. For this we need a variant of the
Chern-Simons theory introduced in Eq.(\ref{lagr}) for the first state introduced in
Eq.(\ref{first}). In this case we have four gauge fields
$a_{i}^{F \sigma}, a_{i}^{B \sigma}; i = 0,1, \sigma = \uparrow, \downarrow$
\begin{eqnarray}
\vec{\nabla} \times \vec{a}^{F \sigma} & = & 2 \pi ( 2 \rho^{F \sigma} +
                                                 2 \rho^{B \sigma}) \nonumber \\
\vec{\nabla} \times \vec{a}^{B \sigma} & = & 2 \pi ( 2 \rho^{F \sigma} +
                                                  \rho^{B \uparrow} + \rho^{B \downarrow}),
\end{eqnarray}
acting on $\Psi_{\sigma}$ and $\Phi_{\sigma}$, fermi and bose fields respectively.
In fact, as can be easily seen, we have only three independent gauge fields
$ a_{c} = \frac{a^{B \uparrow} + a^{B \downarrow}}{2} = \frac{a^{F \uparrow} + a^{F \downarrow}}{2}$,
$ a_{fs} = \frac{a^{B \uparrow} - a^{B \downarrow}}{2}$, and
$ a_{s} = \frac{a^{F \uparrow} - a^{F \downarrow}}{2}$,
and, therefore,
\begin{equation}
\frac{i k a^{fs}_{1}}{2 \pi} = \rho^{F \uparrow} - \rho^{F \downarrow},
\end{equation}
and
\begin{equation}
\frac{i k a^{s}_{1}}{2 \pi} = \rho^{\uparrow} - \rho^{\downarrow}.
\end{equation}
Then, similarly as before, we can write down the effective Lagrangian for the
pseudospin part,
\begin{eqnarray}
{\cal L}_{ps} = & \frac{1}{2 \pi} a_{0}^{s} i k a_{1}^{fs} +
\frac{1}{2 \pi} a_{0}^{fs} i k (a_{1}^{s} - a_{1}^{fs}) - \frac{1}{2} (\frac{k}{2 \pi})^{2}
V_{s} |a_{1}^{s}|^{2} & \nonumber \\
                & + |a^{s}_{0}|^{2} K_{00} + |a^{s}_{1}|^{2} K_{11} + \delta \rho_{s} B_{0} & \nonumber \\
                & + \frac{1}{4} \frac{m \omega^{2}}{\overline{\rho}_{b} k^{2}}
|\delta \rho_{s}^{b}|^{2} - a_{0}^{fs} \delta \rho_{s}^{b} - \frac{\overline{\rho}^{b}}{m}
|a^{fs}_{1}|^{2}.&
\end{eqnarray}
We first integrate out $a_{0}^{fs}$ which gives us the expected constraint on the
pseudospin density of bosons, $\delta \rho_{s}^{B}$. Using this constraint and then
integrating out $a_{1}^{fs}$ and $a_{1}^{s}$ we get for the pseudospin density - density
response the following expression,
\begin{equation}
{\cal L}^{eff} = \frac{\frac{1}{2} (\frac{k}{2 \pi})^{2} B_{0}^{2}}{ W^{4} +
(\frac{k}{2 \pi})^{2} V_{s} - 2 K_{11} - \frac{1}{2} \frac{m \omega^{2}}{\overline{\rho}_{b} (2 \pi)^{2}}},
\end{equation}
where
\begin{equation}
W^{4} = - \frac{\{\frac{1}{2} \frac{m \omega^{2}}{\overline{\rho}_{b}^{2} (2 \pi)^{2}} \}^{2}}{ (\frac{k}{2 \pi})^{2} \frac{1}{2 K_{00}}
-\frac{1}{2} \frac{m \omega^{2}}{\overline{\rho}_{b} (2 \pi)^{2}} + \frac{2 \overline{\rho}^{b}}{m}}.
\end{equation}
First in the limit $\omega \rightarrow 0$ (and then $k \rightarrow 0$) we see that
the system is compressible. Second, taking into account that
\begin{equation}
K_{11} \approx -\frac{k^{2}}{12 \pi m} + i \frac{2 \overline{\rho}_{f}}{k_{F}} \frac{\omega}{k},
\end{equation}
we see that the pseudospin gapless mode, in the case of the (111) state simply
$\omega^{o} = \sqrt{\frac{2 V_{s}}{m} \overline{\rho}_{b}} k$, does not exist as an
eigenmode in this case. It is nearly so if we have in mind that the fraction of
bosons in this state is to be considered small, and so is the imaginary part of
$K_{11}$ when for the frequency we consider the one that takes to zero the real part
of the denominator in ${\cal L}^{eff}$.

In fact also for the state in Eq.(\ref{first}) we find that there is a pseudospin
eigenmode with a dispersion relation $ \omega = \omega^{o}(k) + i c k^{3}$ where $c$
is a constant. Therefore it is slightly dissipative what we do not expect from the
Goldstone mode \cite{moo}. In the following section we will consider the variational
constructions, Eq.(\ref{first}) and Eq.(\ref{second}), with $p$ - pairing of
composite fermions, introduced in Ref. \cite{ste} , that cures the dissipation problem of the
construction in Eq.(\ref{first}) and leads, as we will show, to
incompressible behavior also in the pseudospin channel of the state in Eq.(\ref{second}).

\section{Composite fermion pairing and a possibility for a pseudospin liquid}
In Ref. \cite{ste} $p$-wave CF pairing was proposed as a way of lowering ground state energy
of the state in Eq.(\ref{first}). That was explicitly shown on a basis of
Monte-Carlo calculations in which the paired states are excellent variational
ansatzes with respect to the true ground states. In the context of the
phenomenological Chern-Simons theories based on the proposed wave functions
it is not hard to show that in the case of $p$-wave CFs pairing the state in
Eq.(\ref{first}) acquires a pseudospin mode without an imaginary term which
must be then a Goldstone mode  predicted by
the theory of the ordered state for small $d$ \cite{moo}.

The calculation begins by noting that $p$-wave pairing of CFs is simply a condensation
into "11-1" state by the way of the Cauchy identity,
\begin{eqnarray}
&\prod_{i<j}(w_{i \uparrow} - w_{j \uparrow})
\prod_{k<l}(w_{k \downarrow} - w_{l \downarrow})
\prod_{p,q}(w_{p \uparrow} - w_{q \downarrow})^{-1} & = \nonumber \\
&\det\{\frac{1}{w_{p \uparrow} - w_{q \downarrow}}\}.&.
\end{eqnarray}
In fact, as stated in Ref.\cite{ste}  , the $p$-pairing they found is with the pairing
function $ g(w) \sim \frac{1}{w^{*}}$, i.e. a $p_{x} - i p_{y}$ instead of
$p_{x} + i p_{y}$  pairing, and in the context of the Chern-Simons theory that
means that the gauge fields on CFs acquire negative sign, so that in the end we have
\begin{equation}
\frac{ik}{2 \pi} a_{1}^{F \sigma} = \frac{ik}{2 \pi} a_{1}^{B \sigma} =
\rho_{\sigma}^{F} + \rho_{-\sigma}^{F} + \rho_{\sigma}^{B} + \rho_{-\sigma}^{B}.
\end{equation}
Then it is easy to show, similarly as before, by integrating out $\delta \rho_{s}^{F}$ and
$\delta \rho_{s}^{B}$, in the pseudospin channel, that the Goldstone mode has
dispersion $ \omega(k) = \sqrt{\frac{2 V_{s}}{m} (\overline{\rho}_{b} + \overline{\rho}_{f})} \cdot k$.
We get the same result if we consider $p_{x} + i p_{y}$ pairing instead of
$p_{x} - i p_{y}$.

On the other hand if we consider the state in Eq.(\ref{second}) with
fermi type intercorrelations between the fermi and bose part and
introduce the $p_{x} - i p_{y}$ pairing between CFs, in the CS language we have
\begin{equation}
\frac{ik}{2 \pi} a_{1}^{F \sigma} =
\rho_{\sigma}^{F} + \rho_{-\sigma}^{F} + 2 \rho_{\sigma}^{B},
\end{equation}
and
\begin{equation}
\frac{ik}{2 \pi} a_{1}^{B \sigma} =
\rho_{\sigma}^{B} + \rho_{-\sigma}^{B} + 2 \rho_{\sigma}^{F},
\end{equation}
i.e. $\frac{ik}{2 \pi} a_{1}^{fs} = \rho_{1}^{F} - \rho_{2}^{F} $
and $\frac{ik}{2 \pi} a_{1}^{s} = \rho_{1}^{B} - \rho_{2}^{B} $. Then
the pseudospin part of the effective Lagrangian is
\begin{eqnarray}
{\cal L}_{ps}^{eff} = & \frac{1}{2 \pi} a_{0}^{s} i k a_{1}^{fs} +
\frac{1}{2 \pi} a_{0}^{fs} i k a_{1}^{s} & \nonumber \\
& + \frac{1}{4} \frac{m \omega^{2}}{\overline{\rho}_{B} k^{2}}
|\delta \rho_{s}^{B}|^{2} - a_{0}^{fs} \delta \rho_{s}^{B} - \frac{\overline{\rho}^{B}}{m}
|a^{fs}_{1}|^{2}& \nonumber \\
& + \frac{1}{4} \frac{m \omega^{2}}{\overline{\rho}_{F} k^{2}}
|\delta \rho_{s}^{F}|^{2} - a_{0}^{s} \delta \rho_{s}^{F} - \frac{\overline{\rho}^{F}}{m}
|a^{s}_{1}|^{2}& \nonumber \\
&- \frac{1}{2} (\frac{k}{2 \pi})^{2}
V_{s} |a_{1}^{s} + a_{1}^{fs}|^{2}  - \frac{ik}{2 \pi} (a_{1}^{s} + a_{1}^{fs}) B_{o}.&
\end{eqnarray}
In few steps, by reducing $ {\cal L}_{ps}^{eff}$ into an effective, quadratic
expression in $B_{o}$ we can find out the density-density correlator. It
vanishes in the $k \rightarrow 0$ limit and signals that the state in Eq.(\ref{second})
in which CFs pair is an {\em incompressible} state. It is our expectation that in this
state exist four kinds of merons, characteristic also to the ordered
pseudospin state. Therefore the state, if a ground state of an electron system,
should represent a quantum Hall i.e., in the low-energy limit, a topological
phase.

Even in existing numerical work we find a support for our expectation.
In Ref. \cite{par} , in the data that represent the excitation spectrum of the
bilayer $\nu = 1$ system at $ d = 1.5 l_{B}$ ($l_{B}$ is the
magnetic length), i.e. in the transition region, obtained
by exact diagonalization on a torus, we can see signatures of the
nearby topological phase. Namely, in the low-lying spectrum dissociated
from the Goldstone mode excitations, exists a fourfold degenerate energy level that
we expect represents expected four ground states on the torus of the
topological theory \cite{fre}. (We do not interpret these states as ordered spiral
states as in Ref.\cite{par} because no nearby, low-lying excitations can be seen in the
existing data.) Because of the importance of the intralayer correlations
in the transition region, the mixed state with fermi type intercorrelations
between the bose and fermi part should compete with the true ground state
and is very relevant to the physics of the region. The topological theory
in question is $ U_{2}(1) \otimes U_{2}(1)$ because it supports two kinds of
semionic quasiparticles \cite{fre}. (Meron fractional statistics is semionic.)
On the other hand the counterflow experiments \cite{kel,tut} do not support a ``perfect"
superfluid scenario with a Kosterlitz-Thouless transition. Instead the data
on the counterflow longitudinal resistance show an activated behavior, for a range
of higher temeratures,
very similar to the usual data of a quantum Hall phase.
Therefore it might be that due to an increased importance of the intralayer
correlations in the counterflow experiments, the topological phase
stabilizes and with it a gapped behavior even in Hall
 resistance \cite{kel,tut}. Because of the non-chiral flow of
the currents in the experiment the relevant topological theory should be
$ U_{2}(1) \otimes \overline{U_{2}(1)}$ double non-chiral pseudospin
liquid. The theory is invariant under combined time reversal and $Z_{2}$,
an exchange of layer indecies, operations implied by the experimental setup.
Furthermore it is well-known that the zero-bias peak in the tunneling conductance
is not as of the usual Josephson effect in superconductors due to dissipation.
The reason for this should also be found in the physics of the state in
Eq.(\ref{second}) that incorporates the effect of the increased intralayer
correlations in the transition region in which the experiment occured.

\section{Discussion: Phase diagram from the wavefunctional approach}
When we take into account what we found out in previous sections, in the
scope of the wavefunctional approach, the phase diagram, in the absence
of disorder, may well have an intermediate phase
between the superfluid phase and the two decoupled Fermi-liquid like phases.
The wavefunctional approach tells us (sections III and IV) that in the
intermediate phase the pseudospin stiffness may go to zero (meron deconfinement)
but the density of composite bosons stays finite. The density of
bosons disappears at the transition to the two decoupled CF Fermi seas.
\begin{figure}
\centering
\vspace{3mm}
\includegraphics[width=0.9\linewidth]{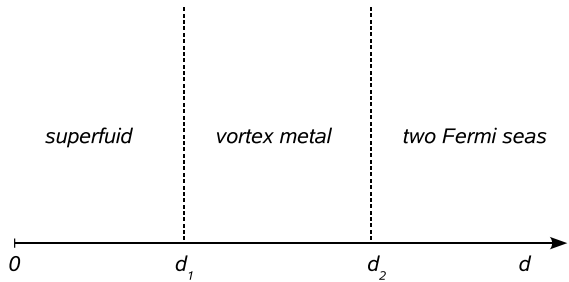}
\caption{The possible phase diagram of the bilayer as a function of
the distance between layers in the presence of disorder
at zero temperature.}\label{four}
\end{figure}

This scenario of the influence of quantum disordering in a superfluid is
often described as a result of creation of vortex-antivortex pairs (loops)
that cause phase fluctuations. The phase fluctuations may cause the
disappearance of the superfluid long range order, but may leave the
ordering amplitude (boson density) nonzero. In the Appendix
we will argue that creation of (2 + 2) CFs in the (111) condensate of
CBs as in the construction in Eq.(\ref{first}) can be viewed as
creation of two closely spaced paires of merons. Each pair consists of
two merons of the same vorticity but opposite charge. On the other hand
the creation of (2 + 2) CFs in the condensate of CBs as in the
construction in Eq.(\ref{second}) leads to meron pairs that are not
closely spaced inside each pair. (The construction in Eq.(\ref{second})
suits the increasing intralayer correlations with $d$.) This motivates
a picture of the intermediate phase as either a condensate of closed meron-antimeron loops of any size - a pseudospin liquid
\cite{fre} (section IV) or a soup of dissociated
meron-antimeron pairs (loops) i.e. a vortex metal (section III)
.
The presence of disorder  may stabilize
the vortex metal phase as in Fig. 4, or maybe a phase separated version \cite{fer}.

\section{Conclusion}
By systematically investigating possible candidates for ground state
wave functions for the bilayer $\nu = 1$ system we reached the conclusion that the
ordered state that supports a Goldstone mode in the pseudospin channel for general
$d$ can be described as a mixture of CBs and $p$-paired CFs with dominant
bose (111) type intercorrelations between the two components. Because of
the increasing intralayer correlations with $d$, a state
that competes with the ordered one can be described as a mixture of CBs and
$p$ - paired CFs with dominant fermi type intercorrelations between
two components. This state that describes a topological phase with four
gapped meron quasiparticles may
cause (already observed) activated behavior (decaying exponentially with
a gap) behavior of the counterflow longitudinal and Hall resistances at higher temperatures.
At lower temperatures, because of the presence of disorder, a compressible
in the pseudospin channel
version of this state, a vortex metal, may come out and cause the observed phenomenology of
the bilayer ``imperfect superfluid".

There is a vast literature on the bilayer. We are planning a
separate publication on the features of the vortex metal phase and
comparisons
 to the previous, experimental and theoretical 
 work. Here we will mention only theoretical work that like ours
stresses the importance of the interactions and quantum disordering.
In Ref.\cite{ye} an excitonic square lattice solid phase was proposed.
Here we discussed only translationally invariant, homogenous possibilities
 and inhomogenous states, Ref.\cite{ye}, may also come as candidates.
(The inhomogenous, phase separated version of the mixed CB-CF states
was first proposed in Ref.\cite{sh} before the homogenous kind
\cite{srm}.) In Ref.\cite{do} interactions only drive a phase
transition into a state with algebraic (quasi-)long-range superfluid
order. The numerics, without impurities and not biased by boundary
conditions and special number of electrons to the Wigner crystal
formation, Ref.\cite{par}, tells us that only the true superfluid
order exists in the pertinent experimental region. Therefore we want
to stress again what our work suggests. Quantum disordered states
(``double pseudospin liquid") are nearby the true superfluid ground
states for distance between the layers pertinent to the experiments.
We need also impurities to stabilize a version of these disordered
states, ``vortex metal state", at lower temperatures. Interactions
caused quantum disordering is necessary but not sufficient condition
for the explanation of the dissipative phenomenology of the
experiments.

The author would like to thank the Kavli Institute for Theoretical Physics
for its hospitality during the workshop on ``Topological Quantum Computation",
where part of this work was completed.
She also thanks M. Levin, N. Read, Z. Te\v{s}anovi\'{c}, and X.-G. Wen for discussions.
The work was supported by
Grant No. 141035  of the Serbian Ministry of Science.

\appendix
\section{}
In this Appendix we will give arguments that the four (2 + 2) CF inclusion
into the (111) state, i.e.
\begin{eqnarray}
&{\cal S}_{\uparrow} \{ (z_{1 \uparrow} - z_{2 \uparrow})
(\exp\{i\vec{k}\vec{z}_{1 \downarrow}\} - \exp\{i\vec{k}\vec{z}_{2 \downarrow}\})\}
& \times \nonumber \\
&{\cal S}_{\downarrow} \{ (z_{1 \downarrow} - z_{2 \downarrow})
(\exp\{i\vec{k}\vec{z}_{1 \downarrow}\} - \exp\{i\vec{k}\vec{z}_{2 \downarrow}\})\}&
\Psi_{111}, \nonumber \\
& &
\label{twotwo}
\end{eqnarray}
where ${\cal S_{\uparrow}}$ and ${\cal S_{\downarrow}}$ are symmetrizers inside
each layer, and we omitted for the sake of simplicity the overall projection
to the LLL, corresponds to two meron pairs, of opposite (up and down)
vorticity. Inside each pair merons have the same vorticity but opposite charge.
Therefore the two meron constructions that can act on the (111) state and
produce such a pair is (see Eq.(\ref{mcon}))
\begin{eqnarray}
\hat{S}^{\uparrow}(w_{1}) \hat{S}^{\uparrow}(w_{2}) \equiv
&\prod_{i} \frac{z_{i \uparrow} - w_{1}}{|z_{i \uparrow} - w_{1}|} \exp\{ - \sum_{i}
\frac{C}{2 |z_{i \uparrow} - w_{1}|} \} &  \times \nonumber \\
&\prod_{j} \frac{z_{j \uparrow} - w_{2}}{|z_{j \uparrow} - w_{2}|} \exp\{ \sum_{j}
\frac{C}{2 |z_{j \uparrow} - w_{2}|} \} & , \nonumber \\
& &
\end{eqnarray}
and analogously for the down pair. These expressions are effective i.e.
valid in the long distance approximation. Nevertheless in this Appendix
we will take that they are qualitatively correct even for shorter
distances. Then in the quasiparticle (``fractional statistics") representation \cite{lau}
the wave function that describes the two merons in Eq.(\ref{twotwo}) is
\begin{equation}
\Psi(w_{1}, w_{2}) = (w_{1} - w_{2})^{- \frac{1}{2}} f(w_{1}, w_{2}),
\end{equation}
where due to the mutual semionic statistics between quasiparticles
we have the difference, $(w_{1} - w_{2})$, to the power $- \frac{1}{2}$, and
$f(w_{1}, w_{2})$ is a symmetric function of coordinates, in our case:
\begin{equation}
f(w_{1}, w_{2}) = (\exp\{i \vec{k} \vec{w}_{1}\} - \exp\{i \vec{k} \vec{w}_{2}\})
\frac{(\vec{w}_{1} - \vec{w}_{2})}{|\vec{w}_{1} - \vec{w}_{2}|}.
\end{equation}
There are no Gaussian factors because, when calculated, the interaction
of a meron with a positive background of the corresponding plasma,
\begin{equation}
\int d^{2}z \frac{1}{|z - w|} \exp\{- \frac{|z|^{2}}{2}\},
\end{equation}
is a bounded function of $w$, and when expontiated gives a factor that
weakly depends on $w$.

To get the wave function in terms of electronic coordinates we have to
calculate \cite{lau}
\begin{eqnarray}
&\int d^{2} w_{1} \int d^{2} w_{2} \frac{1}{|w_{1} - w_{2}|} f(w_{1}, w_{2})
\hat{S}^{\uparrow}(w_{1}) \hat{S}^{\uparrow}(w_{2}) & \times \nonumber \\
&\int d^{2} w_{3} \int d^{2} w_{4} \frac{1}{|w_{3} - w_{4}|} f(w_{3}, w_{4})
\hat{S}^{\downarrow}(w_{3}) \hat{S}^{\downarrow}(w_{4}) & \Psi_{111}. \nonumber \\
&     &
\label{elwf}
\end{eqnarray}
The combined exponentials in the two meron construction can be expanded
as in the following,
\begin{eqnarray}
&\exp\{ \sum_{i} \frac{C}{2 |z_{i \uparrow} - w_{2}|} \}
\exp\{ - \sum_{i} \frac{C}{2 |z_{i \uparrow} - w_{1}|} \} = & \nonumber \\
& 1 + \frac{C}{2} \sum_{i} \frac{(\vec{z}_{i} - \vec{w}_{1}) \vec{\delta}}{
(\vec{z}_{i} - \vec{w}_{1})^{2} |\vec{z}_{i} - \vec{w}_{1}|}&  \nonumber \\
& - \frac{C}{2} \sum_{i} \{ \frac{\vec{\delta}^{2}}{
(\vec{z}_{i} - \vec{w}_{1})^{2} |\vec{z}_{i} - \vec{w}_{1}|} + 3
\frac{(\vec{z}_{i} - \vec{w}_{1}) \vec{\delta}}{
(\vec{z}_{i} - \vec{w}_{1})^{4} |\vec{z}_{i} - \vec{w}_{1}|} \} & \nonumber \\
&+ \frac{1}{2!} (\frac{C}{2})^{2} \sum_{i}
\frac{ \{(\vec{z}_{i} - \vec{w}_{1}) \vec{\delta}\}^{2}}{(\vec{z}_{i} - \vec{w}_{1})^{6}}& \nonumber \\
&+ \frac{1}{2!} (\frac{C}{2})^{2} \sum_{i \neq j}
\frac{(\vec{z}_{i} - \vec{w}_{1}) \vec{\delta} \cdot (\vec{z}_{j} - \vec{w}_{1}) \vec{\delta}}{
(\vec{z}_{i} - \vec{w}_{1})^{2} |\vec{z}_{i} - \vec{w}_{1}|
(\vec{z}_{j} - \vec{w}_{1})^{2} |\vec{z}_{j} - \vec{w}_{1}|}& \nonumber \\
& + o(|\delta|^{3}) &
\label{expan}
\end{eqnarray}
where $\vec{\delta} = \vec{w}_{2} - \vec{w}_{1}$, and for the sake
of clarity we supressed the up arrow. To the same order of accuracy we
can rewrite the above expression as
\begin{eqnarray}
& 1 + \frac{C}{2} \sum_{i} \frac{(\vec{z}_{i} - \vec{w}_{2}) \vec{\delta}}{
(\vec{z}_{i} - \vec{w}_{2})^{2} |\vec{z}_{i} - \vec{w}_{2}|}&  \nonumber \\
&+ \frac{1}{2!} (\frac{C}{2})^{2} \sum_{i}
\frac{(\vec{z}_{i} - \vec{w}_{1}) \vec{\delta} \cdot
(\vec{z}_{i} - \vec{w}_{2}) \vec{\delta}}{
(\vec{z}_{i} - \vec{w}_{2})^{2} |\vec{z}_{i} - \vec{w}_{1}|
(\vec{z}_{i} - \vec{w}_{1})^{2} |\vec{z}_{i} - \vec{w}_{2}|}& \nonumber \\
&+ \frac{1}{2!} (\frac{C}{2})^{2} \sum_{i \neq j}
\frac{(\vec{z}_{i} - \vec{w}_{1}) \vec{\delta} \cdot
(\vec{z}_{j} - \vec{w}_{2}) \vec{\delta}}{
(\vec{z}_{i} - \vec{w}_{2})^{2} |\vec{z}_{i} - \vec{w}_{1}|
(\vec{z}_{j} - \vec{w}_{1})^{2} |\vec{z}_{j} - \vec{w}_{2}|}.&
\label{expend}
\end{eqnarray}
The contribution of the first two terms must be negligible or zero
after the integration over $\vec{w}_{1}$ and $\vec{w}_{2}$ which
brings averages of random phases with respect to electron distributions
encoded in $\Psi_{111}$. The contribution of the third term, after
picking sigularities at $z_{i}$, is identical to zero, but the
contribution of the fourth term, after picking up singularities
at $z_{i \uparrow}$ and $z_{j \uparrow}$ is
\begin{eqnarray}
& f(z_{i \uparrow}, z_{j \uparrow}) \cdot |z_{i \uparrow} - z_{j \uparrow}|& \times \nonumber \\
& \prod_{k \neq i} \frac{(z_{k \uparrow} - z_{i \uparrow})}{|z_{k \uparrow} - z_{i \uparrow}|}
\prod_{l \neq j} \frac{(z_{l \uparrow} - z_{j \uparrow})}{|z_{l \uparrow} - z_{j \uparrow}|}. &
\end{eqnarray}
for the up part of the construction in Eq.(\ref{elwf}).
We get a similar contribution for the down construction, and
therefore, up to some phase factors, we get Eq.(\ref{twotwo}).
The phase factors are there to ensure the right flux count through the system,
the problem we neglected by writing our wave functions
(Eq.(\ref{first}) and Eq.(\ref{second})) in the thermodynamic limit.

By examining (2 + 2) CF construction in which Fermi-type
intercorrelations are dominant, i.e. the one as the construction in
Eq.(\ref{second}) for (2 + 2) CFs, we can find out that the small
distance approximation as in Eq.(\ref{expend}) is not applicable.
Moreover its form suggests that much higher order terms of the
expansion in Eq.(\ref{expan}) are relevant, and
therefore two merons are well separated in this construction.

\end{document}